\documentstyle[prl,aps,epsf,multicol]{revtex}
\begin{document}
\draft
\tighten

\title{Transversely Driven Charge Density Waves and Striped Phases 
of High-T$_c$ Superconductors: The Current Effect
Transistor}

\author{Leo Radzihovsky}
\address{Physics Department, University of Colorado, Boulder, CO 80309--0390}
\author{John Toner} 
\address{Dept. of Physics,
Materials Science Inst., and Inst. of Theoretical Science,
University of Oregon, Eugene, OR 97403}

\date{\today}
\maketitle

\begin{abstract}
We show that a normal (single particle) current density $J_x$ {\em
transverse} to the ordering wavevector $2k_F{\bf\hat{z}}$ of a charge
density wave (CDW) has dramatic effects both above and {\em below} the
CDW depinning transition. It exponentially (in $J_x$) enhances CDW
correlations, and exponentially suppresses the longitudinal depinning
field. The intermediate longitudinal I-V relation also changes,
acquiring a {\em linear} regime. We propose a novel ``current effect
transistor'' whose CDW channel is turned on by a transverse
current. Our results also have important implications for the recently
proposed ``striped phase'' of the high-T$_c$ superconductors.
%  
%
%We study charge density waves (CDW's) in the presence of a normal
%(single-particle) current density $J_x$ {\em transverse} to the
%ordering wavevector $2{k_F}{\bf\hat{z}}$ below and above the CDW
%longitudinal depinning transition. We demonstrate that even for
%stationary CDW's (below the depinning transition) the transverse
%current radically alters the long distance correlations. This has many
%dramatic consequences for experiments: above a ``critical'' {\em
%transverse} normal current $J_c$, the CDW correlation length grows
%exponentially with $J_x$ as $\xi_L(J_x>J_c)=\xi_L\ e^{(J_x-J_c)/J_c}$
%and concomitantly the CDW depinning electric field decays
%exponentially with $J_x$. For $J_x>J_c$, the high {\em longitudinal}
%current I-V is qualitatively modified, exhibiting an intermediate {\em
%linear} regime.  We propose a novel ``Current Effect Transistor'', in
%which the CDW channel transport is turned on by a transverse normal
%current.
%
\end{abstract}

%\twocolumn

\begin{multicols}{2}
%\narrowtext

While the equilibrium properties of charge density waves\cite{FLR} (as
well as other disordered periodic media) are by now fairly well
understood, recent attention has focussed on their far less well
understood {\em non-equilibrium} dynamical
properties.\cite{SCF}-\cite{BMR}

All classical treatments of CDW's to date have included only the local
displacement $u({\bf r},t)$ of the CDW, which is related to the
electron density via $\rho({\bf r},t)=\rho_0 + \rho_1 \cos(2 k_F( z +
u({\bf r},t))$, as the only important hydrodynamic mode of the
problem.

In this Letter we will argue that while such an approach is correct
for the statics, the total electron density $\rho$ is crucial for the
correct description of the {\em non-equilibrium} dynamics.  We will
present a dynamical model that incorporates this additional mode and
explore its consequences.

This additional mode allows current to flow even when the CDW itself
is stationary. The effects are particularly dramatic when the current
flows {\em perpendicular} to ${\bf q_0}=2k_F{\bf\hat{z}}$, the
ordering wavevector of the CDW. Once this transverse current $J_x$
exceeds a {\em crossover} value $J_c$, it makes the CDW much more
ordered. The correlation length of a three dimensional (3d) CDW along
the direction ($\bf\hat x$) of the transverse current obeys
\begin{equation}
\xi_x(J_x)\approx\left\{\begin{array}{lr}
\xi_L\;, &\mbox{for}\; J_x<J_c\;,\\
\xi_L {J_c\over J_x}e^{2(J_x-J_c)/J_c}\;, &\mbox{for}\; J_x>J_c\;,
\end{array} \right.
\label{xi_x}
\end {equation}
where $\xi_L$ is the ``Larkin length'' (see below) of the CDW in zero
transverse current, which is finite due to the random pinning of the
CDW by impurities.

For directions perpendicular to the transverse current the correlation
lengths also grow exponentially with $J_x$, but with {\em precisely
half} the growth rate:
\begin{equation}
\xi_{\perp}(J_x)\approx\xi_L\left({J_c\over J_x}\right)
e^{(J_x-J_c)/J_c}\;,\;\;
\mbox{for}\;\; J_x>J_c\;,
\label{xi_perp}
\end{equation}

The ``critical'' current $J_c$ in the above expressions is:
\begin{equation}
J_c=\sigma_\infty E_T(0)(k_F\xi_L)(\rho_n/\rho_{cdw})\times O(1)
\label{Jc_physical}
\end{equation}
where $E_T(J_x=0)$ is the threshold electric field along $2k_F{\bf
\hat{z}}$ necessary to depin the CDW, and
$\sigma_\infty$ the conductivity in that direction for $E_z\gg
E_T(0)$. For typical NbSe$_3$ samples $J_c\sim 10^3-10^4$ Amps/cm$^2$.
These quite high values can be greatly reduced by increasing disorder
(lowering $\xi_L$).

In addition to being directly observable by X-ray scattering
measurements of the CDW correlation length, these exponential
dependences of the correlation lengths on $J_x$ also have striking
consequences for the I-V characteristics along $\bf\hat{z}$. If we
apply a ``longitudinal'' electric field $E_z$ along $\bf\hat z$ while
maintaining a fixed $J_x$, we find the (zero temperature) threshold
depinning field $E_T(J_x)$ required to first make the CDW move is
given by
\begin{equation}
E_T(J_x)=\left\{\begin{array}{lr} E_T(0)\;, &\mbox{for}\;
J_x<J_c\;,\\ 
E_T(0)\left({J_x\over
J_c}\right)^{3/2}e^{-{2(J_x-J_c)/J_c}}\;, &\mbox{for}\;
J_x>J_c\;.
\end{array} \right.
\label{E_T}
\end {equation}
Correspondingly, the shape of the high velocity part of the I-V is
also drastically modified.  We find that for longitudinal fields $E_z$
in the range, $E_T(J_x)<<E_z<< E_T(0)(J_x/J_c)^2$, the I-V is {\em
Ohmic} (i.e. linear) with the longitudinal conductivity given by
$\sigma_z(J_x)=\sigma_\infty\left(1-(J_c/J_x)\times O(1)\right)$.  For
the largest fields $E_z>>E_T(0)(J_x/J_c)^2$, $J_z(E_z)$ crosses over
to the $J_x=0$ result: $J_z(E_z)=\sigma_\infty
E_z\left(1-c\sqrt{E_T(0)/E_z}\right)$.\cite{SCF} All of our results
for the longitudinal I-V characteristics in the presence of a
transverse current are summarized in Fig.1.

Our results suggest a novel ``current effect transistor'' (CET), in
which the threshold field {\em along} $\bf q_0$ is controlled by a
current driven {\em transversely} to $\bf q_0$.

All of the above results should also apply to the recently proposed
``striped phase'' of the high-T$_c$ cuprate oxide
superconductors,\cite{tranquada} since the macroscopic symmetries of
that system (i.e., unidirectional, lattice-incommensurate modulation
of the charge density) are the same as those of a CDW, as is the
ubiquitous presence of disorder.\cite{RTunpublished}
\begin{figure}[bth]
{\centering
\setlength{\unitlength}{1mm}
\begin{picture}(150,70)(0,0)
\put(-3,-25){\begin{picture}(150,70)(0,0) 
\includegraphics{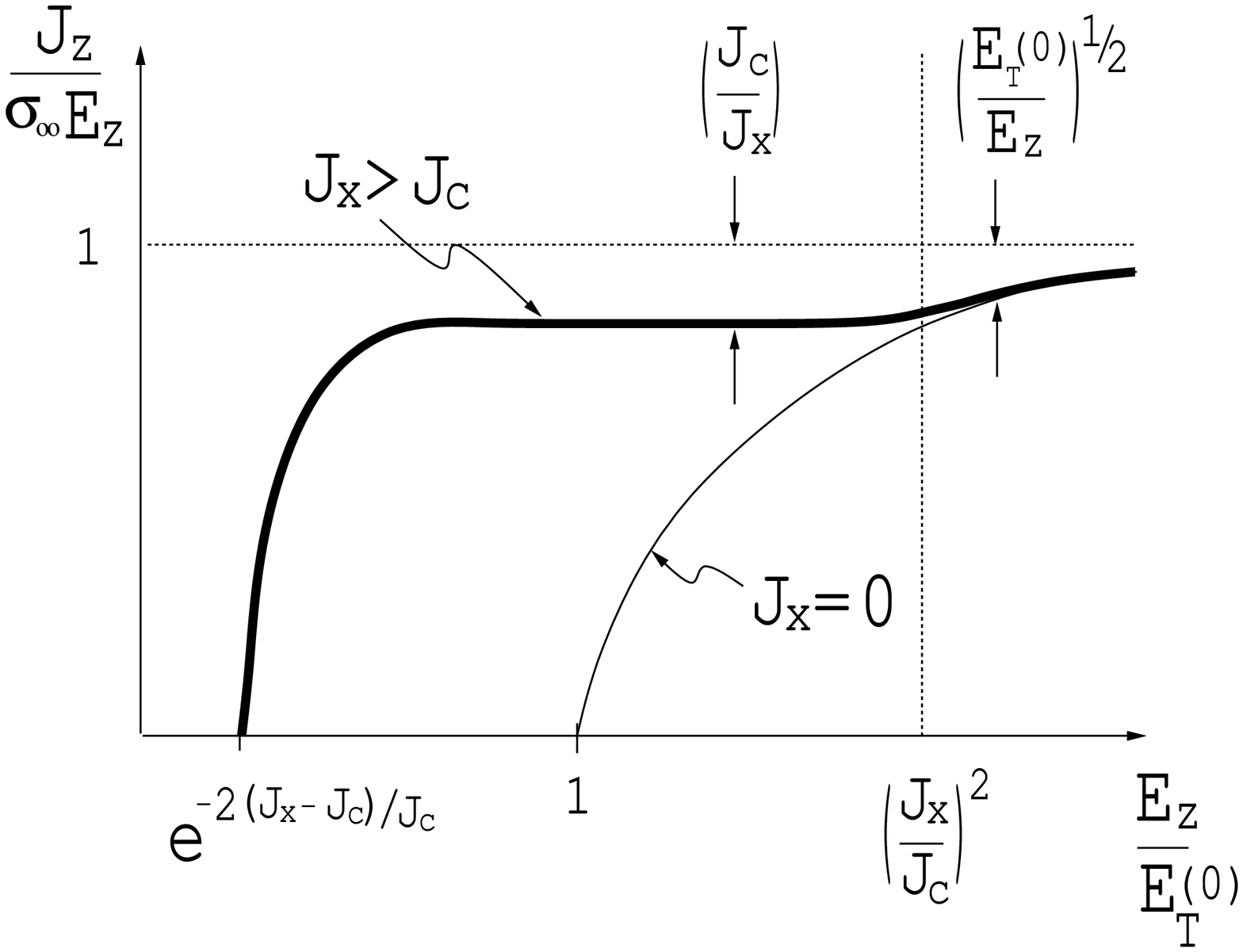}
\end{picture}}
\end{picture}}
Fig.1:{(a) Schematic I-V curve illustrating new intermediate Ohmic
high velocity regime that exists for transverse currents $J_x>J_c$.}
\label{iv}
\end{figure}

We now present our model and derive these results.  General principles
dictate that there are two types of long wavelength hydrodynamic
variables: Goldstone modes associated with broken symmetries, and
conserved variables. In CDW's, the phonon mode $u$ associated with the
breaking of translational symmetry is the sole Goldstone mode, while
the total electron density $\rho$ is a conserved variable. Contrary to
the naive expectation, $\rho$ is {\it not} determined solely by the
compression of the CDW (i.e. $\delta\rho\neq -\rho_{cdw}\partial_z
u$). This is in direct analogy with ordinary 3d crystals, for which
the density of vacancies and interstitials (or equivalently the total
particle density) must be included in the proper hydrodynamic
description\cite{Martin}.

Although the total charge density $\rho$ is a conserved field, in
CDW's, because of long-range Coulomb interactions\cite{Kamenev}, it
does {\em not} constitute a {\em slow} hydrodynamic mode, unlike
neutral systems, e.g. the transverse smectic phase of a driven vortex
lattice\cite{BMR}.
%Instead, $\rho$ is a fast mode, responding
%at the plasma frequency $\omega_p=(4\pi\rho_0 e^2/m)^{1/2}$. On time
%scales longer than $\omega_p^{-1}$ it does not modify the
%long-wavelength {\em equilibrium} CDW dynamics.
%\cite{finiteK}

In contrast, in {\em driven} CDW's one {\em must} consider total
charge density dynamics to even obtain the correct equation of motion,
to which we now turn.

We consider a pinned CDW with an ordering wavevector ${\bf q_0}=2k_F{\bf\hat
z}$ and explore the consequences of a {\em ``normal''} current ${\bf J}$.  Our
main conceptual point is that, in contrast to all previous treatments such a
current makes even the {\em stationary} CDW a highly {\em non-equilibrium}
system in which the ${\bf\hat J}\rightarrow -{\bf\hat J}$ symmetry is broken. 
Lacking the spatial inversion symmetry along the direction of such a
single-particle current, general symmetry principles therefore dictate that
the equation of motion for $u({\bf r},t)$ {\em must} admit and therefore {\em
will} contain non-equilibrium terms that break this inversion symmetry.  The
most important of these is the ``convective'' term ${\bbox{\tilde
v}}\cdot{\bbox\nabla} u$.  The striking effects of this should be
experimentally observable.

We explore these in a transverse geometry, in which the charge current
${\bf J}$ flows {\em transversely} to the CDW ordering wavevector
$2k_F{\bf\hat z}$, i.e. ${\bf J}=J_x {\bf\hat{x}}$. The hydrodynamics
of a CDW in the presence of such transverse current $J_x$ is described
by
\begin{equation}
  \gamma (\partial_t + {\bbox{\tilde{v}}}\cdot{\bbox{\nabla}})u({\bf
  r},t) = K\nabla^2 u + F({\bf r},u) + e \rho_{cdw} E_z\;,
\label{cdw_u2}
\end{equation}
where $E_z$ is the externally imposed electric field along ${\bf k}_F$
and $\rho_{cdw}$ is the average CDW electron number density.  We take
the quenched random force $F({\bf r},u)$ in Eq.\ref{cdw_u2} to be
Gaussian distributed with zero mean and $\overline{ F({\bf r},u)
F({\bf r}',u')} = \Delta(u-u') \delta({\bf r - r'})$, where the
overbar denotes a disorder average. The function $\Delta(u)$ is
periodic with the CDW lattice spacing, $a=2\pi/2k_F$.

Because Galilean symmetry is broken by the underlying lattice and
quenched disorder the ``convective'' coupling $\bbox{\tilde{v}}$ is
different from (but probably on the order of) the ``normal'' electron
velocity ${\bbox v}={\bf J}/\rho_n e$, where $\rho_n$ is the
``normal'' electron density.  For simplicity of notation we have taken
$K_x=K_y=K_z=K$, however all of our results can be trivially extended
to anisotropic elasticity.

The essential difference between our model Eq.\ref{cdw_u2} and the
commonly used Fukuyama-Lee-Rice (FLR)\cite{FLR} model is the
``convective'' $\tilde{v}_x\partial_x u$ term, which physically arises
because a tipped ($\partial_x u\neq 0$) ``layer'' of the CDW deflects
the transverse normal current downward by an angle
$\theta\sim\partial_x u$, leading to a reaction force back on the CDW,
proportional to $\partial_x u$ and the normal current $J_x$.

This ``convective'' term leads to a velocity-dependent crossover
length scale $L_c(\tilde{v}_x)=K/(\gamma\tilde{v}_x)$. On length
scales $L_x$ (along x) $< L_c$, the convective term is unimportant and
the CDW behaves as though it is in equilibrium;\cite{comment} for
$L_x>L_c$, non-equilibrium effects set in.

In spatial dimensions $d<4$, even arbitrarily weak pinning destroys
the translational order of the CDW for length scales longer than the
so-called Larkin length scale $\xi_L$.\cite{FLR,Larkin} On this length
scale the disordering effect of the random pinning on $u$ just
balances the ``elastic'' forces that try to keep the CDW ordered.

Clearly, for vanishing or small transverse currents (small
$\tilde{v}_x$) the standard FLR-Larkin (FLRL) calculation of $\xi_L$,
which ignores $\tilde{v}_x$, applies, giving in 3d $\xi_L= K^2
a^2/\Delta(0)$.

Given the two length scales $\xi_L$ and $L_c(\tilde{v}_x)$, one can
define a crossover transverse velocity $\tilde{v}_c$ by
$\xi_L=L_c(\tilde{v}_c)$, giving
$\tilde{v}_c={K/(\gamma\xi_L)}={\Delta(0)/(\gamma K a^2)}$.  Taking
$\tilde{v}_x\approx J_x/(\rho_n e)$, we can obtain a crossover {\em
current} density $J_c=\rho_n e K/(\gamma\xi_L)$.  Using the standard
FLR result for the longitudinal threshold electric field $E_T=K
a/(\rho_{cdw} e \xi_L^2)$ and the large-field limit of the
longitudinal CDW conductivity
$\sigma_{\infty}=e^2\rho_{cdw}^2/\gamma$, we obtain
Eq.\ref{Jc_physical}, with $J_c$ expressed in terms of experimental
observables.  For $J_x<J_c$, the equilibrium FLRL calculation of
$\xi_L$ is valid and the CDW translational correlation length is
independent of the transverse current $J_x$.

Interesting transverse current dependent phenomena occur for
$J_x>J_c$. In this regime, we can calculate the new CDW correlation
length $\xi_\perp(\tilde{v}_x)$ in the $\perp$-direction by asking how
big in the direction perpendicular to $x$ the system would have to be
before the mean-squared fluctuations $\overline{\langle u^2\rangle}$
in the position of the CDW exceed $a^2$. For a finite system of width
$L_\perp$, in $d<4$ dimensions, the mean-square phonon fluctuations
$\overline{\langle u^2({\bf r})\rangle}= \int_{q_\perp>L_\perp^{-1}} d
q_x d^{d-1} q_\perp\overline{\langle |u({\bf q})|^2\rangle}$ can be
calculated by spatially Fourier transforming Eq.\ref{cdw_u2} and
looking for the static solution for the spatial Fourier transform
field $u({\bf q})$.

For a system with $L_\perp\leq\xi_\perp(\tilde{v}_x)$, the $u$
fluctuations remain smaller than the lattice spacing $a$, and so we
can replace $F({\bf r},u)$ with its $u=0$ value. The static equation
of motion then reads
\begin{equation}
(i\gamma \tilde{v}_x q_x + K q^2) u({\bf q})=F({\bf q};u=0)\;,
\label{u_q}
\end{equation}
where $F({\bf q};u=0)$ is the spatial Fourier transform of $F({\bf
r};u=0)$. Solving this for $u({\bf q})$, and then computing
$\overline{\langle |u({\bf q})|^2\rangle}$ from the result, we obtain
\begin{equation}
\overline{\langle |u({\bf q})|^2\rangle}={\Delta(0)\over \gamma^2
\tilde{v}_x^2 q_x^2 + K^2 q^4}\;,
\label{u_q2}
\end{equation}
Inserting this into expression for $\overline{\langle u({\bf
r})^2\rangle}$ and defining $\xi_\perp(\tilde{v}_x)$ as the value of
$L_\perp$ at which $\overline{\langle u({\bf r})^2\rangle}=a^2$ we
obtain an implicit equation for $\xi_\perp(\tilde{v}_x)$
\begin{equation}
(\xi_L/\xi_\perp(\tilde{v}_x))^{4-d}=
f\left(\xi_\perp(\tilde{v}_x)/L_c(\tilde{v}_x)\right)
\end{equation}
where $f(y)=2(4-d)\int^\infty_1 dx x^{d-2}/(g_+ g_-^{1/2}+g_+^{1/2}
g_-)$, with $g_\pm(x,y)=x^2+y^2/2\pm y(x^2+y^2/4)^{1/2}$.

Evaluating the integral in $d=3$ in the $J_x<<J_c$ and $J_x>>J_c$
limits and using $\tilde{v}_x\approx J_x/(\rho_n e)$ gives
Eq.\ref{xi_perp} for $\xi_\perp(J_x)$, which matches the $J_x=0$
Larkin length at $J_x=J_c$, i.e., $\xi_\perp(J_c)\approx \xi_L$. The
correlation length $\xi_x(J_x)$ along the transverse current $J_x$
($\tilde{v}_x$) is easily computed from $\xi_\perp(J_x)$ to be given
by Eq.\ref{xi_x}.

Physically, the exponential dependence of the CDW correlation length
for $J_x>J_c$ arises due the strong suppression of the CDW roughness
by the transversely moving normal charge carriers, which ``stiffen''
the CDW elasticity by momentum transfer with it. The transverse
convective term lowers the upper critical dimension (below which the
random pinning disorders the CDW) from
$d_{uc}(\tilde{v}_x<\tilde{v}_c)=4$ to
$d_{uc}(\tilde{v}_x>\tilde{v}_c)=3$.

The strong exponential dependence of the Larkin length on a transverse
single particle current $J_x$ has a number of striking consequences
which should be experimentally observable (aside from being directly
probed in X-ray scattering). It is easy to show that at low
temperatures and weak disorder, the threshold {\em longitudinal} field
(along the ordering wavevector $2k_F{\bf \hat{z}}$) is given by
$E_T(J_x)=E_T(0)\left({V_L(0)/V_L(J_x)}\right)^{1/2}$, where $E_T(0)=K
a/(e\rho_{cdw}\xi_L^2)$ is the $J_x=0$ longitudinal threshold electric
field and $V_L(J_x)$ is the transverse current dependent Larkin
volume.  This decay of the longitudinal threshold field $E_T$ with
Larkin volume is physically associated with the $\sqrt N$ statistic
reduction in the strength of the random pinning force, when averaged
over the correlation volume $V(J_x)=\xi_\perp^2(J_x)\xi_x(J_x)$. Using
our results for $\xi_{x,\perp}(J_x)$, Eq.\ref{xi_x},\ref{xi_perp} in
the expression for $E_T(J_x)$ above, we obtain Eq.\ref{E_T}.

We now turn to the effects of the transverse current $J_x$ for $E_z$
well above the longitudinal threshold field $E_T(J_x)$. Perturbing
about the uniformly moving CDW state $u_0=v_z t$, we seek a solution
to Eq.\ref{cdw_u2} of the form $u\left({\bf r},t \right) = v_z t +
\delta u \left({\bf r},t \right)$, where $v_z$ is the mean velocity of
the CDW, to be determined self-consistently, and $\delta u
\left({\bf r},t
\right)$ a zero mean fluctuation (assumed small) about this uniformly
moving state.  Inserting this Ansatz into Eq.\ref{cdw_u2}, expanding
for small $\delta u$, and averaging over the disorder, we have
$v_z=\gamma^{-1}e\rho_{cdw} E_z + \delta v_z$, with
\begin{equation}
\gamma\delta v_z\approx\left.\overline{\partial_u F({\bf r},u)\delta u({\bf
r},t)} \right|_{u=v_z t}\;,
\label{v_z}
\end {equation}
where we have used the facts that $\overline{F({\bf r},v_z t)} =
\overline{\nabla^2\delta u}=0$.  We can obtain $\partial_u F\delta u$
from the equation of motion for $\delta u({\bf r},t)$. To leading
order in $\delta u$, that equation reads:
\begin{equation}
(\gamma\partial_t + \gamma{\bbox{\tilde{v}}}\cdot{\bbox{\nabla}}
-K\nabla^2)\delta u =\sum_{q_n} F({\bf r},q_n)\,e^{iq_n v_z t}\;,
\label{delta_u_EOM}
\end {equation}
where we have used the fact that $F({\bf r},u)$ is a periodic function
of $u$ and $q_n=n q_0$, with $n\in Z$.

Solving Eq.\ref{delta_u_EOM} for $\delta u$ and using it inside
Eq.\ref{v_z} for $\delta v_z$ we obtain
\begin{equation}
\gamma\delta v_z\approx-\sum_{q_n}\int{d^d q\over (2\pi)^d}
{\Delta_{q_n}q_n(v_z q_n-\tilde{\bbox v}\cdot{\bbox q})\over
\gamma^2(v_z q_n-\tilde{\bbox v}\cdot{\bbox q})^2+(K q^2)^2}
\label{v_z2}
\end{equation}
We can verify a posteriori that the wavevector integral in this
expression is dominated by $q_z \ll q_n$. This together with
$\tilde{\bbox v}\sim {\bbox v}$ implies $\tilde{\bbox v}\cdot {\bf
q}\approx \tilde{v}_x q_x$, so:
\begin{equation}
\gamma\delta v_z\approx-\int{d^d q\over (2\pi)^d}
{2\Delta_0\tilde{v}_x q_0(q_1-q_x)\over
\gamma^2\tilde{v}_x^2(q_1-q_x)^2+K^2 q^4}
\label{v_z3}
\end{equation}
where we have defined $q_1\equiv q_0 v_z/\tilde{v}_x$ and kept only
$n=1$; higher random force harmonics ($n>1$) do not qualitatively
change the results.

We can asymptotically evaluate the integral in Eq.\ref{v_z3} in the
limits: (i) $q_x^*\ll q_1$ and (ii) $q_x^*\gg q_1$, where $q_x^*$ is
the characteristic value of $q_x$ that dominates the integral. A
simple analysis shows that in (i) $q_x^*=\sqrt{\gamma v_z q_0/K}$. The
condition $q_x^*<q_1$ then leads to the constraint $E_z/E_T(0) > \mbox
{Max}\{1,(J_x/J_c)^2\}$. Ignoring $q_x$ relative to $q_1$ in
Eq.\ref{v_z3} we find that the SCF result\cite{SCF} holds at these
highest longitudinal fields: $\delta v_z/v_z =
-(E_T(0)/E_z)^{(4-d)/2}\times O(1)$. For $J_x<J_c$, this is the only
regime.

For $J_x>J_c$, we find that a new regime (regime (ii)) exists, when
the longitudinal fields $E_z/E_T(0) < (J_x/J_c)^2$. Here $q_1$ is
subdominant to $q_x$ and the integral above can be evaluated by a
Taylor expansion in $q_1$. Changing variables $q_x-q_1\rightarrow q_x$
and expanding in $q_1$ we find
\begin{equation}
{\delta v_z\over v_z}\approx-8\Delta_0 q_0^2 K^2 
\int {d^d q\over(2\pi)^d} {q^2_x q^2
\over \left[\gamma^2 \tilde{v}_x^2 q^2_x + 
K^2 q^4\right]^2}
\label{v_z4}
\end{equation}

The integral on the right hand side is {\em finite} in the infra-red
and the ultraviolet for $1<d<4$. All higher terms in this $q_1$
expansion are likewise finite. Evaluating the integral in $d=3$ we
obtain
\begin{equation}
{\delta v_z\over v_z}\approx-{\Delta q_0^2 \over K\gamma\tilde{v}_x}
\times O(1)=-{\tilde{v}_c\over\tilde{v}_x}\times O(1)
\label{v_z5}
\end{equation}

A more detailed analysis\cite{RTunpublished} shows that the above
expansion breaks down as $E_T(\tilde{v}_x)$ is approached from above.
This can be seen simply by assuming that $\tilde{v}_z$ in
Eq.\ref{cdw_u2} is of order the $v_z$ we found above, and using that
result in a calculation of $\langle\delta u({\bf r},t)^2\rangle$. We
find that this approaches $a^2$, invalidating our small $\delta u$
expansion, for $E_z\sim$ the threshold field $E_T(\tilde{v}_x)$ found
earlier. This breakdown signals the pinning of the CDW for
$E_z<E_T(\tilde{v}_x)$.

This conclusion is consistent with our earlier observation that $J_c$
is a {\em crossover} (as opposed to {\em critical}) transverse
current, and with similar results for the related transverse smectic
phase of a driven vortex lattice.\cite{BMR} Our results for the
transversely driven CDW I-V characteristics are summarized in Fig.1.

Another observable quantity sensitive to transverse current is the
Fourier transform $I({\bf r})$ of the structure function:
\begin{eqnarray}
\overline{\langle\rho^*_{2k_F}({\bf r})\rho_{2k_F}(0)\rangle}
=\left\{\begin{array}{lr} e^{-r/\xi_L}\;, &\mbox{for}\;J_x<J_c\;,\nonumber\\
r_\perp^{-\eta(J_x)} g_\eta\left({r_\perp^2\over x\xi_L}{J_x\over
J_c}\right)\;, &\mbox{for}\; J_x\gg J_c\;,
\end{array} \right.\nonumber
%\label{I_r}
\end{eqnarray}
where $\eta(J_x)=O(1)\times J_c/J_x$ and $g_\eta(z)$ is a scaling
function readily calculated to lowest order in weak disorder.  It is
important to keep in mind that the expression above is only valid for
length scales smaller than the Larkin length.

On scales larger than the Larkin length, the small $u$ expansion of
the random force is invalid. However, using the functional
renormalization group developed in Ref.\cite{BMR}, we can show that
$u$ fluctuations are logarithmic for both $J_x<J_c$, and for
$J_x>J_c$.  In the former case, the phonon correlation function
$C(r_\perp, x)$ (which grows linearly with $r$ for $r<\xi_L$) first
(at the Larkin scale) crosses over to the familiar super-universal
(for $2<d<4$) logarithmic dependence on $r$\cite{VillainFernandez}.
At scales larger than $L_c(\tilde{v}_x)$ the correlation function is
also logarithmic (but with non-universal coefficient), now as a
consequence of the fact that non-equilibrium effects have lowered the
upper-critical dimension to $d_{uc}=3$.\cite{BMR} For $J_x>J_c$ the
intermediate regime is absent and there is a direct crossover at the
Larkin scale to the logarithmic growth associated with non-equilibrium
effects.

%An important point of principle is that our non-equilibrium model
%Eq.\ref{cdw_u2} differs from the equilibrium FLR model even for
%stationary CDW (i.e., for $E_z<E_T$) with no transverse current. This
%is because even in this situation $\tilde{v}_z$ in Eq.\ref{cdw_u2} is
%still nonzero (since it is allowed by symmetry) any time $E_z\neq
%0$. One must therefore reexamine this problem for $E_z<E_T$, to
%determine, e.g., if these non-equilibrium effects make the Larkin
%length scale $E_z$ dependent, which might, in turn, change the
%threshold field. Fortunately for FLR, a simple analysis of
%Eq.\ref{cdw_u2} shows that a purely longitudinal field $E_z$ only
%changes the Larkin length $\xi_L$ for $E_z>E_z^c\propto 1/\xi_L$, while
%the threshold field $E_T\propto 1/\xi_L^2$. Hence, for weak pinning,
%where $\xi_L\rightarrow\infty$, the CDW always depins {\em before}
%non-equilibrium effects become important. We therefore conclude that
%in a {\em purely longitudinal} geometry the permeation mode, below the
%depinning transition, does not drastically modify results obtained
%from models that ignored it.

Finally we note that our results, in particular Eq.\ref{E_T}, might
have an interesting device application: the ``current effect
transistor'' (CET). Imagine a channel made of e.g.,
NbSe$_3$.\cite{Thorne} At low $T$, and longitudinal fields below the
threshold field $E_T$, such a channel is insulating. For $E_z>E_T$ the
CDW depins and the channel becomes highly conducting. Our analysis
demonstrates that this {\em longitudinal} threshold field $E_T(J_x)$
is a strong function of the {\em transverse} current $J_x$ for
$J_x>J_c$, as summarized by Eq.\ref{E_T}. One can therefore turn on
the CDW transport by a transverse current $J_x$ which depins the CDW
by depressing the longitudinal threshold field $E_T(J_x)\sim
e^{-2(J_x-J_c)/J_c}$.  Though theoretically interesting, this
``device'' would almost certainly be too slow and low gain to be
practical.

In conclusion, we have argued that previous treatments of
sub-threshold CDW dynamics have ignored an important ``permeation''
(total charge density) mode. We demonstrate that this mode radically
changes the CDW dynamics in the transverse current geometry.

L.R. acknowledges discussions with A.C. Neto and support by the NSF
DMR-9625111 and the A.P. Sloan Foundation; J.T. was supported by the
NSF DMR-9634596.

\end{multicols}

\end{document}